\documentclass{article}
\usepackage{ dsfont }
\usepackage{dcase2020_techrep,amsmath,graphicx,url,times,booktabs, tabularx}
\usepackage{xcolor}

\usepackage{hyperref}

\title{Device-Robust  Acoustic Scene Classification Based on Two-Stage Categorization And Data Augmentation}

\name{Hu Hu$^{1}$,
      Chao-Han Huck Yang$^{1}$,
      Xianjun Xia$^{2}$,
      Xue Bai$^{3}$,
      Xin Tang$^{3}$,
      }
\secondlinename{	
      Yajian Wang$^{3}$,
      Shutong Niu$^{3}$,
      Li Chai$^{3}$,
      Juanjuan Li$^{2}$,
      Hongning Zhu$^{2}$,
      Feng Bao$^{2}$,
      }
\thirdlinename{
      Yuanjun Zhao$^{2}$,
      Sabato Marco Siniscalchi$^{1, 4}$,
      Yannan Wang$^{2}$,
      Jun Du$^{3}$,
      Chin-Hui Lee$^{1}$
    }
\address{$^1$School of Electrical and Computer Engineering, Georgia Institute of Technology, GA, USA \\
$^2$Tencent Media Lab, Tencent Corporation, China\\
$^3$University of Science and Technology of China, HeFei, China\\
$^4$Computer Engineering School, University of Enna Kore, Italy
}

\begin{document}

\ninept
\maketitle

\begin{sloppy}

\begin{abstract}
In this technical report, we present a joint effort of four groups, namely GT, USTC, Tencent, and UKE, to tackle Task 1 -  Acoustic Scene Classification (ASC) in the DCASE 2020 Challenge. Task 1 comprises two different sub-tasks: (i) Task 1a focuses on ASC of audio signals recorded with multiple (real and simulated) devices into ten different fine-grained classes, and (ii) Task 1b concerns with classification of data into three higher-level classes using low-complexity solutions. For Task 1a, we propose a novel two-stage ASC system leveraging upon ad-hoc score combination of two convolutional neural networks (CNNs), classifying the acoustic input according to three classes, and then ten classes, respectively. Four different CNN-based architectures are explored to implement the two-stage classifiers, and several data augmentation techniques are also investigated. For Task 1b, we leverage upon a quantization method to reduce the complexity of two of our top-accuracy three-classes CNN-based architectures. On Task 1a development data set, an ASC accuracy of 76.9\% is attained using our best single classifier and data augmentation. An accuracy of 81.9\% is then attained by a final model fusion of our two-stage ASC classifiers. On Task 1b development data set, we achieve an accuracy of 96.7\% with a model size smaller than 500KB \footnote{\scriptsize Code available:  \url{https://github.com/MihawkHu/DCASE2020_task1}}.

\end{abstract}


\section{Introduction}
\label{sec:intro}
Acoustic scene classification (ASC) is concerned with correctly identifying real-world sounds into a set of given environment classes, such as metro station, street traffic, or public square. An acoustic scene sound contains much information and rich content that makes accurate scene prediction difficult, and thereby an intriguing research problem. ASC has thus been an attractive research field for decades, and the Detection and Classification of Acoustic Scenes and Events (DCASE) challenge \cite{dcase2016, dcase2017, dcase2018, dcase2020} provides the benchmark data and a competitive platform to promote sound scene research and analyses.

In DCASE 2020, there are two different sub-tasks of Task 1. Task 1a focuses on the robustness problem of ASC system. Its goal is to promote research to solve the device mismatch issue, which is a common case in ASC applications. The key aim is to design a device-invariant system, which can classify ten scene audios recorded by different devices well, without leveraging any device information in the evaluation stage. Task 1b focuses on the model size of the ASC system. The goal is to build a three-class classifier occupying no more than 500KB.

We describe our submitted systems for the two sub-tasks of DCASE 2020. For Task 1a, we build a two-stage acoustic scene classification system, which includes a three-class classifier and a ten-class classifier. The final predicted class is based on the score fusion of these two classifiers. Four different convolutional neural network (CNN) based models are used in our two-stage classifier. Moreover, several data augmentation strategies are adopted to reduce the device dependency of our models. Model ensemble of 4 CNN-based systems provided a significant boost of the ASC performance. For task 1b, we build a small-size model at first, and then use quantization method to compress the well-trained model. A model can be compressed to $1/8$ of the original size by this way. In our experiments, the ensemble of two smaller models can get better evaluation results than a single model, when they have the same level model size.

\section{Acoustic Scene Classification System}
\label{sec:sys}
\subsection{Two-Stage Classification Procedure}
For Task 1a, we build a two-stage ASC system, which includes two different classifiers and outputs the class of the input audio scene choosing among ten classes. As shown in figure~\ref{fig:system}, the first classifier is a three-class classifier, and it classifies an input audio scene into one of three main classes, including in-door, out-door, and transportation. The second classifier is a ten-class classifier, which classifies a given input audio scene into one of ten basic scene classes, including airport, shopping mall, metro station, pedestrian street, public square, street traffic, tram, bus, metro, park. For each input audio, its final scene class is chosen by score fusion of those two classifiers. If we let $\mathds{C}^1$ and $\mathds{C}^2$  denote the set of three main classes and ten classes, respectively, and let $F^1$ and $F^2$  indicate the output of the first and second classifier, respectively. The final predicted class $Class(x)$ for the input  $x$ is:
\begin{equation}
    Class(x) = \underset{q, (p\in \mathds{C}^1, q\in \mathds{C}^2, p \supset q)}{\operatorname{argmax}}\ F^1_p(x) * F^2_{q}(x), \nonumber
\end{equation}
\noindent where $p \supset q$ means that $p$ can be though of a super set of $q$. For example, transportation class is the super set for bus, tram, and metro classes. Therefore, the probability of an input audio clip to be from the public square scene is equal to the product  of the probability of out-door place given by $F^1_p(x)$ and the probability of public square given by $F^2_q(x)$.

\begin{figure}[t]
  \centering
  \includegraphics[width=0.8\linewidth]{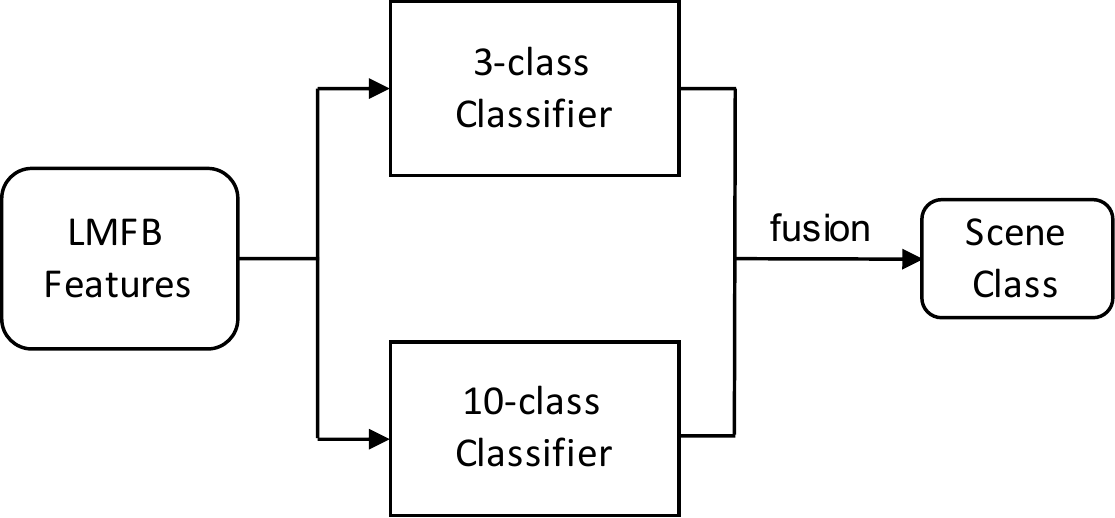}
  \caption{The proposed two-stage acoustic scene classification system for Task 1a.}
  \label{fig:system}
  \vspace{-4mm}
\end{figure}

\subsection{CNN-based Classifiers}
Five CNN based architectures, which differ one from another for specific details concerning the usage of (i) time and frequency pooling (sub-sampling) operations, (ii) independent frequency sub-band analysis,  (iii) shortcut connections (i.e., residual mapping functions), and (iv) number of convolutional layers, are used:

\begin{itemize}
    \item FCNN (fully convolutional neural network): FCNN is a VGG \cite{vgg}-like fully convolutional network based on 9 stacked convolutional layers with small kernel sizes. Each convolutional layer is followed by a Batch Normalization operation and ReLU activation layer. Dropout is also used in the convolutional layers from five to the last in order to alleviate over-fitting issues.  A $2 \times 2$ max-pooling layer is appended after the second, fourth, and eighth ReLu-based layers. Channel attention is applied to each output channel of the last convolutional layer, followed by a global pooling layer. Finally, a 10-way softmax layer is used to generate the final classification decision. In Task 1b, we use an FCNN with a similar architecture but much fewer parameters, and we refer to that model as small-FCNN.

    \item fsFCNN (frequency sub-sampling controlled fully convolutional neural network): Through our experiments on DCASE 2020 Task 1a, we noticed that reducing max-pooling in the frequency axis helped contrasting overfitting issues; thereby, we deployed a neural architecture very similar to the above-mentioned FCNN but having 11 stacked convolutional layers. Moreover, $2 \times 2$ max-pooling layer is appended only after the second and fourth ReLU-based layers. A $1 \times 2$ max-pooling layer instead follows the sixth, and eighth  ReLU-based layers. Channel attention is applied to each output channel of the last convolutional layer, followed by a global pooling layer. Finally, a 10-way softmax layer is used to generate the final classification result.

    \item fsFCNN-s (frequency sub-sampling controlled fully convolutional neural network with split frequency bands): Each input feature map is first split ('-s' in the model name) into two sub-feature maps along a frequency dimension. If there are $N$ frequency bins, frequency bins between 0 and $(N/2)-1$ are processed by an fsFCNN, and frequency bins between $N/2$ and $N$ would be independently processed by another fcFCNN. The processing will happen in parallel up to the ReLU-based layer. Then the two processed streams would be concatenated and processed by two further convolutional layers. Finally,  a global pooling layer and 10-way softmax is used to get the final scene classification decision.

    \item Resnet (17-layer residual network): Resnet model is a residual network \cite{resnet}. We use the network structure proposed in  \cite{dcase-resnet}, which has 17 convolutional layers. There is no frequency sub-sampling throughout the whole network. Each input feature map is divided into two sub-feature maps along the frequency dimension. To be specific, if we have $N$ frequency bins, the first $N/2$ and the second half are processed by two parallel stacked convolutional layers. Thus, we have a two-stage model structure. Like FCNN, a global pooling layer and 10-way softmax are used to get the final utterance level prediction results. Different from the structure in \cite{dcase-resnet}, in our final submission, we double the filter number of each convolutional layer, and we denote it to Resnet-d.

    \item Mobnet (MobileNet-v2): Mobnet is based on MobileNet-v2 \cite{mobnet}. The key feature of Mobnet is its low complexities despite the high accuracies that can be attained, as demonstrated in \cite{mobnet}. Mobnet uses lightweight depth-wise convolutions to process features in the intermediate expansion layer. We leveraged a relatively small-size Mobnet to tackle Task 1b.
\end{itemize}
\begin{table*}[ht]
\centering
\caption{Experimental results on Task 1a. 'Scale' means scaling input data to [0,1]. 'Spec-aug' means performing a mini-batch level spectrum augment method. '5*Aug' means using extra five kinds of augmented data, including spectrum correction, pitch shift, speed change, add random noise, and mix audios. 'Two-stage' means using a two-stage classification system. All 'Y' in the table means we used that method.}
\label{tab:task1a}
\vspace{0.2cm}
\begin{tabular}{l|l|ccccc|cccc|c|c}
\hline
\hline
Idx. & System    & Scale & Spec-aug & 5*Aug & \begin{tabular}[c]{@{}c@{}}Reverb\\ +DRC\end{tabular} & Two-stage & \begin{tabular}[c]{@{}c@{}}A \\ acc. \%\end{tabular} & \begin{tabular}[c]{@{}c@{}}B\&C\\ acc. \%\end{tabular} & \begin{tabular}[c]{@{}c@{}}s1-s3\\ acc. \%\end{tabular} & \begin{tabular}[c]{@{}c@{}}s4-s6\\ acc. \%\end{tabular} & \begin{tabular}[c]{@{}c@{}}val\\ loss\end{tabular} & \begin{tabular}[c]{@{}c@{}}Avg \\ acc. \%\end{tabular} \\
\hline
\hline
(0)   & Baseline \cite{dcase2020}   & -     & -        & -     & -       & -       & 70.6                                          & 61.6                                             & 53.3                                             & 44.3                                               & 1.37                                               & 54.1                                                      \\
\hline
(1)   & Resnet   & -     & -        & -     & -       & -       & 78.5                                            & 72.6                                              & 68.8                                               & 68.1                                               & 1.28                                               & 70.2                                                       \\
(2)   & Resnet   & Y     & -        & -     & -       & -       & 78.8                                            & 72.1                                              & 69.3                                               & 69.5                                               & 1.30                                               & 71.0                                                       \\
(3)   & Resnet   & Y     & Y        & -     & -       & -       & 80.3                                            & 73.5                                              & 71.4                                               & 67.7                                               & 1.26                                               & 71.6                                                       \\
(4)   & Resnet-d & Y     & Y        & -     & -       & -       & 79.7                                            & 74.8                                              & 71.9                                               & 70.7                                               & 1.28                                               & 73.0                                                       \\
(5)   & Resnet-d & Y     & Y        & Y     & Y       & -       & 83.0                                            & 76.1                                              & 73.6                                               & 71.0                                               & 1.23                                               & 74.6                                                       \\
\hline
(6)   & FCNN     & Y     & Y        & Y     & -       & -       & 88.5                                            & 79.8                                              & 74.6                                               & 70.6                                               & 1.08                                               & 76.0                                                       \\
(7)   & FCNN     & Y     & Y        & Y     & Y       & -       & 87.3                                            & 79.5                                              & 75.7                                               & 73.0                                               & 1.07                                               & 76.9                                                       \\
\hline
(8)   & fsFCNN   & Y     & Y        & Y     & Y       & -       &    83.9                                            & 78.6                                              & 75.4                                              & 72.8                                               & 1.13                                               & 76.2                                       \\
\hline
(9)  & (7)+(8)      & Y     & Y        & Y     & Y       & -       & 87.6                                            & 79.8                                              & 76.4                                               & 73.8                                               & 1.12                                               & 77.5                                                       \\
(10)  & (5)+(7)+(8)  & Y     & Y        & Y     & Y       & -       & 87.0                                            & 81.5                                              & 78.0                                               & 76.9                                               & 1.22                                               & 79.4                                                       \\
\hline
(11)  & (7)+(8)       & Y     & Y        & Y     & Y       & Y       & 89.1                                            & 82.4                                              & 78.5                                               & 77.5                                               & 0.84                                               & 80.2                                                       \\
(12)  & (5)+(7)+(8)  & Y     & Y        & Y     & Y       & Y       & 87.9                                            & 84.1                                              & 80.4                                               & 79.9                                               & 0.94                                               & 81.9          \\
\hline
\hline
\end{tabular}
\end{table*}

\section{Data Augmentation Strategy}
\label{sec:dataaug}

A key element of our submission is the use of several data augmentation strategies. The first four strategies discussed below do not generate extra training data; whereas, the remaining  schemes generate extra training data. In Task 1a, all the following data augmentation methods except channel confusion were used. In Task 1b, mixup, data augmentation schemes are limited to channel confusion, and spectrum augment.

\begin{itemize}
    \item Mixup: It was proposed in \cite{mixup} and is often adopted to train ASC models. We use mixup with alpha equal to $0.4$. Mixup is performed at a mini-batch level: Two data batches, along with corresponding labels, are randomly mixed in each training step.
    \item Random cropping: It was proposed in  \cite{dcase-resnet}. During the training stage, the input data is randomly cropped into a fixed-length along the time axis. In our experiments, the input data with the size of $423 \times 128 \times 3$ is cropped into $400 \times 128 \times 3$ input feature map. Due to quantization constraints on dynamic ranging, random cropping is not applied to Task 1b.
    \item Channel confusion: It is used in Task 1b only. Two channels in input data are randomly swapped.
    \item Spectrum augmentation: It was proposed in \cite{spec-aug} and showed a significant boost of the automatic speech recognition performance. In our implementation, we carried out spectrum augmentation over each input feature map. We applied it at a mini-batch level. For a batch data in the training step, each feature map is randomly masked in both time and frequency axes. With respect to \cite{spec-aug} terminology, we set the parameter of time and frequency mask to 10\% of their dimensions, respectively.
    \item Spectrum correction: It was proposed in \cite{spec-corr} and demonstrated moderate device adaptation properties. However, spectrum correction aims at transforming a given input spectrum to that of a reference, possibly ideal, device. Different from the original idea, we here employ spectrum correction as a data augmentation technique. To this end, we had to modify the original procedure as follows: (i)  we create a reference device spectrum, by averaging the spectrum from all training devices except that from device A; (ii)  we correct the spectrum of each training waveform collected with device A to obtain extra data.
    \item Reverberation + Dynamic Range Compression (DRC): It's inspired to the procedure used by the organizers to generated simulated devices, namely s1-s11. In fact, s1-s11 data is generated by adding reverberation followed by DRC to audio collected with device A. 
    \item Pitch shift: For each training waveform, we randomly shift the pitch based on the uniform distribution.
    \item Speed change: For each training waveform, we randomly change the audio speed based on the uniform distribution. If the output waveform is longer than the original one,  extra samples are dropped from the end. If shorter, padding is applied till attaining the same input length.
    \item  Random noise: For each training waveform, random Gaussian noise is added.
    \item Mix audios: We randomly mix two audios from the same acoustic scene class. It's device-independent; this data augmentation scheme might help simulate a new "device."
\end{itemize}

\section{Quantization for Model Compression}
\label{sec:quant}
\vspace{-0.1cm}

In Task 1b, the main goal is to keep the system size within 500 Kilobytes (KB). A post-training quantization method, which is provided by Tensorflow2 \cite{tensorflow}, is used to compress our neural models. Quantization not only reduces the model size but also improves hardware accelerator latency with little degradation in final classification accuracy. In our submission, we used dynamic range quantization (DRQ). In DRQ, neural weights are quantized from floating-point to integer having a 8-bit precision. Leveraging DRQ, we thus transferred our neural architectures from a 32-bit TensorFlow format to a 8-bit TensorFlow-lite format, which compresses the model size to about $1/8$ of its original size. According to our experimental evidence, such a compression method resulted in a minor ASC classification drop.

\vspace{-0.2cm}
\section{Experimental Setup \& Results}
\label{sec:exp}
\vspace{-0.1cm}

\subsection{Feature Extraction}
DCASE2020 ASC audio clips have a fixed-length of 10 seconds. Log-mel filter bank (LMFB) features were used in our experiments as audio features. The input audio waveform is analyzed with a  $2048$ SFFT points,  a window size of $2048$ samples, and a frameshift of $1024$ samples. The librosa~\cite{mcfee2015librosa} library is used to generate LMFBs, and the HTK formula definition for the Mel scale is adopted. Due to different sampling rates for Tasks 1a and 1b, the final spectrogram has 431-time bins in Task 1a and 469-time bins in Task 1b, but the number of frequency bins is 128 in both tasks. Log-mel deltas and delta-deltas without padding were also computed, which reduced the number of time samples to 423 for Task 1a, and 461 for Tasks 1b.  The final input tensor size is thus $423 \times 128 \times 3$ for Task 1a, and $461 \times 128 \times 6$ for Task 1b. Before feeding the speech feature tensors into CNN classifier, we scaled each feature value into [0,1]. 

\subsection{Model Training}
\vspace{-0.1cm}
For the train-test split, we adopt the official recommended way to split the development material. For Task 1a, there are 13965 train audio clips, and 2970 test audio clips. The training set includes audio from devices A, B, C, and s1-s3. The test set covers data from those six devices and extra data from unseen devices s4, s5, and s6. Device A data dominates the training set, which has over 10K utterances. In the test set, the number of waveforms from each device is the same. For Task 1b, there are 9185 train waveforms and 4185 test waveforms. All audio clips are from device A. Stochastic gradient descent (SGD)  with a cosine-decay-restart learning rate scheduler is used to train our models. The maximum and minimum learning rates are 0.1 and 1e-5, respectively. In our final submission, all development data is used. And due to there is no validation data, we use the average output of models when learning rate hits around the minimum number. Keras \cite{keras} is used to implement all our CNN-based models.

\vspace{-0.2cm}
\subsection{Results on Task 1a}
In Table~\ref{tab:task1a}, we report only some of the evaluation results collected on Task 1a due to space constraints. In the training set, device A data accounts for around 75\%, and device B, C, s1-s3 accounts for around 5\%, respectively. Thus, based on the device information of data, we have divided the test set into four different subsets, which represent real source data (device A), real target data (device B \& C), the target seen simulated data (device s1-s3), and target unseen simulated data (device s4-s6).

For comparison, model (0) gives the accuracy of the baseline ASC system provided by the DCASE2020 organizers. For systems (1) to (3), we investigate the effect of scaling and spectrum augmentation. Scaling features to [0,1] give a 0.8\% absolute improvement, and that gain is mainly from device s1-s6. Spectrum augmentation boosts the ASC accuracy from 71.0\% to 71.6\%, yet the accuracy on unseen devices drops significantly. Model (4) doubled Resnet parameters, and that leads to a significant enhancement of the ASC accuracy. Leveraging all data augmentation schemes, we can further improve ASC accuracy up to 74.6\%, which is the best result using the Resnet based models discussed earlier.

Systems (6) to (8) are based on the FCNN architecture. The fsFCNN-s model is not evaluated on this train-test setup, but it's used in our final submission. Moving from Resnet-d to FCNN and leveraging all available training data (original and augmented), ASC accuracy goes from 74.6\% to 76.9\%, which represents the best result with a stand-alone FCNN-based architecture. Comparing models (6) and (7) allows us to appreciate the effect of the reverberation + DRC data augmentation. Indeed, it provides a large improvement in s1-s6 data, especially on unseen s4-s6. That is expected since s1-s6 data is generated by adding reverberation and applying DRC on device A data. Models (7), (8), and fsFCNN-split have very similar structures, yet when  comparing classification results per testing utterance and among those three models,  we observed over 20\% difference in the final decision taken by the three models. Therefore, we keep all of those three models for the model ensemble stage.

Model ensemble is indeed known to boost the ASC accuracy, and we use a simple non-weighted average score fusion as system combination scheme. From the results corresponding to systems (9) to (10), we can see that a two-model combination outperforms any stand-alone systems. By increasing the number of models in the ensemble, we can attain an enhancement of the ASC accuracy from 77.5\% (two models) to 78.1\% (three models). Finally, after model ensemble, a three-class classifier is integrated through a score fusion approach leveraging the two-stage fusion scheme discussed in Section \ref{sec:sys}. From results referring to systems (11) and (12), we can argue that the proposed two-stage fusion strategy significantly improves ASC accuracy. In a two-model ensemble, the ASC accuracy increases from 77.5\% to 80.2\%; and  further increases from 79.4\% to 81.9\% with a three-model ensemble. For the sake of completeness, it should be mentioned that the accuracy of the three-class classification system is 93.2\%.

For our final system submissions, all of the development data is used. Due to resource and time limitations, some of our final submissions are not tested on the above-mentioned train-test split. Our four submissions include: 1) average ensemble of Resnet-d, Resnet-d with attention, fsFCNN-s, and fsFCNN with attention; 2) average ensemble of Resnet-d, FCNN, fsFCNN, and fsFCNN-s trained by different data strategies; 3) average ensemble of all models in 1) and 2); and 4) average ensemble of all models in 1) and 2) except Resnet based models. 

\vspace{-0.2cm}
\subsection{Results on Task 1b}
Table~\ref{tab:task1b} shows our evaluation results on Task 1b. Mobnet and small-fcnn models can attain an ASC accuracy of 95.2\% and 96.4\%, respectively. The model size can be reduced to about $1/8$ of its original size using quantization. For Mobnet, it is observed a 0.4\% performance drop. For small-fcnn,  the performance drop is only equalt to 0.1\%. We thus decide to carry out model ensemble, but we further reduce mobnet and small fcnn parameters before system combination to keep the final combined model size under 500 KB. Therefore, systems in rows 3 and 4 in Table ~\ref{tab:task1b} are slightly different from those in row 1 and 2. In Table~\ref{tab:task1b}, we present the results by logistic regression based model ensemble. As expected, model ensemble achieves better accuracy than a stand-alone system even after quantization.

For our final submission, all of the development data is used. Among our four submissions, the first is from a single small-FCNN model, and the other three are from the ensemble of Mobnet and small-FCNN, including average ensemble of Mobnet and small-FCNN, logistic regression based ensemble of Mobnet and small-FCNN, and two small-FCNN models.

\vspace{-0.3cm}

\begin{table}[h]
\centering
\caption{Experimental results on Task 1b. }
\label{tab:task1b}
\vspace{0.1cm}
\begin{tabular}{l|c|c}
\hline
\hline
Model                                                                    & \begin{tabular}[c]{@{}c@{}}Original\\ acc. \% (size)\end{tabular} & \begin{tabular}[c]{@{}c@{}}Quantization\\ acc. \% (size)\end{tabular} \\
\hline
\hline
Mobnet                                                                   & 95.2 (3.2M)                                                       & 94.8 (411K)                                                           \\
\hline
small-FCNN                                                               & 96.4 (2.8M)                                                       & 96.3 (357K)                                                           \\
\hline
\begin{tabular}[c]{@{}l@{}}Mobnet \\ + small-FCNN-v1\end{tabular}        & 96.8 (1.8M+1.9M)                                                & 96.7 (490K)                                                           \\
\hline
\begin{tabular}[c]{@{}l@{}}small-FCNN-v1 \\ + small-FCNN-v2\end{tabular} & 96.5 (1.9M+2.1M)                                                & 96.3 (499K)                                                          \\
\hline
\hline
\end{tabular}
\end{table}


\vspace{-0.5cm}
\section{Discussion \& Conclusion}
\label{sec:con}
\vspace{-0.1cm}

Although we achieve over 80\% and  95\% ASC accuracy on Task 1a  and Task 1b, respectively, there is yet a large prediction difference between systems with very similar performance. For example, the prediction overlap of FCNN and fsFCNN is only 77\%, which implied that there exists some audio clips are differently predicted by different systems with similar performance. By listening inspection, we realize that several audio clips are difficult to classify even for a human listener because some acoustic scenes are fuzzy by design. To bright ideas:  voices of people in an indoor area can either be from a shopping mall or an airport,  because the same micro-scene, i.e., a small shop, can be in either a shopping mall or an airport. 

In this technical report, we have described our submitted systems to DCASE 2020 Tasks 1a and 1b challenges. For Task 1a, a two-stage classification system is designed, which includes both 3-class classifiers and 10-class classifiers. The final predicted class is determined by the score fusion of these two classifiers. Different fully CNN based classifiers and data augmentation techniques are investigated in our experiments. For Task 1b, small-size FCNN and mobnet are used as a classifier. The quantization method is performed to compress the ASC model to be less than 500KB. From our evaluation of development data, we achieve an accuracy of 81.9\% on Task 1a, and 96.7\% on Task 1b.

\bibliographystyle{IEEEtran}
\bibliography{refs}
%
%
%
%
%
%
%
%
%

\end{sloppy}
\end{document}